\documentclass{article}

\usepackage[preprint]{neurips_2020}

\usepackage[utf8]{inputenc} % allow utf-8 input
\usepackage[T1]{fontenc}    % use 8-bit T1 fonts
\usepackage{hyperref}       % hyperlinks
\usepackage{url}            % simple URL typesetting
\usepackage{booktabs}       % professional-quality tables
\usepackage{amsfonts}       % blackboard math symbols
\usepackage{nicefrac}       % compact symbols for 1/2, etc.
\usepackage{microtype}      % microtypography
\usepackage{amsmath}
\usepackage{amssymb}
\usepackage{mathtools}
\usepackage{booktabs}
\usepackage{multirow}
\usepackage{caption}
\usepackage{subcaption}
\usepackage{graphicx}
\newcommand{\R}{\mathbb{R}}

\DeclareMathOperator*{\argmax}{arg\,max}

\newcommand{\X}{\mathcal{X}}

\title{Active Deception using Factored Interactive POMDPs to Recognize Cyber Attacker's Intent}

\author{%
    Aditya Shinde \\%\thanks{footnote goes here} \\
    Institute for AI\\
    University of Georgia, Athens\\
    GA 30602\\
    \texttt{adityas@uga.edu} \\
   \And
    Prashant Doshi \\
    Institute for AI \& Dept. of Computer Science\\
    University of Georgia, Athens\\
    GA 30602\\
    \texttt{pdoshi@uga.edu} \\
   \And
    Omid Setayeshfar \\
    Dept. of Computer Science \\
    University of Georgia, Athens\\
    GA 30602\\
    \texttt{omid.s@uga.edu} \\
}

\begin{document}

\maketitle

\begin{abstract}
  This paper presents an intelligent and adaptive agent that employs deception to recognize a cyber adversary's intent. Unlike previous approaches to cyber deception, which mainly focus on delaying or confusing the attackers, we focus on engaging with them to learn their intent. We model cyber deception as a sequential decision-making problem in a two-agent context. We introduce factored finitely-nested  interactive POMDPs (I-POMDP$_\X$) and use this framework to model the problem with multiple attacker types. Our approach models cyber attacks on a single honeypot host across multiple phases from the attacker's initial entry to reaching its adversarial objective. The defending I-POMDP$_\X$-based agent uses decoys to engage with the attacker at multiple phases to form increasingly accurate predictions of the attacker’s behavior and intent. The use of I-POMDPs also enables us to model the adversary's mental state and investigate how deception affects their beliefs. Our experiments in both simulation and on a real host show that the I-POMDP$_\X$-based agent performs significantly better at intent recognition than commonly used deception strategies on honeypots.
\end{abstract}

%----------------------------------------------
\section{Introduction}
\label{sec:intro}
%----------------------------------------------

An important augmentation of conventional cyber defense utilizes deception-based cyber defense strategies~\citep{Pingree2018}. These are typically based on the use of decoy systems called {\em honeypots}~\citep{Spitzner2003} with additional monitoring capabilities. Currently, honeypots tend to be passive systems with the purpose of  consuming  the attacker's CPU cycles and time, and possibly logging the attacker's actions. However, the information inferred about the attackers' precise intent and capability is usually minimal. 

On the other hand, honeypots equipped with fine-grained logging abilities offer an opportunity to better understand attackers' intent and capabilities. We may achieve this by engaging and manipulating the attacker to perform actions that reveal his or her true intent. One way of accomplishing this is to employ active deception. Active strategies entail adaptive deception which seeks to influence the attackers' beliefs and manipulates the attackers into performing desired actions~\citep{jajodia2016cyber}. We investigate how multi-agent decision making can be used toward automating adaptive deception strategies to better understand the attacker.

We represent cyber deception on a single host as a decision-making problem between a defender and an attacker. We introduce a factored variant of the well-known interactive partially observable Markov decision process~\citep{gmytrasiewicz2005framework}, labeled as I-POMDP$_\X{}$, to computationally model the decision making of the defender while reasoning about the attacker's beliefs and capabilities as it acts and observes. I-POMDP$_\X{}$ exploits the factored structure of the problem, representing the dynamics and observation function using algebraic decision diagrams, and solving the model using a method that directly operates on these factored representations~\citep{bahar1997algebric}. This brings some level of tractability to an otherwise intractable framework, sufficient to adequately solve the cyber deception domain. I-POMDP$_\X{}$ explicitly models the beliefs of the attacker and the defender throughout the interaction. This allows for detailed inferences about how specific deceptive actions affect the attacker's subjective view of the system. We evaluate the performance of I-POMDP$_\X{}$ in promoting active deception with multiple attacker types both in simulation and on a real host. Our results show that the I-POMDP-based agent learns the intent of the attacker much more accurately compared to baselines that do not engage the attacker or immediately deploy all decoys en masse.

%---------------------------------------------
\section{Background on I-POMDPs}
\label{sec:background}
%---------------------------------------------

Interactive POMDPs (I-POMDPs) are a generalization of POMDPs to sequential decision-making in multi-agent environments~\citep{gmytrasiewicz2005framework,Doshi12:Decision}. Formally, an I-POMDP for agent $i$ in an environment with one other agent $j$ is defined as,
\[
\text{I-POMDP}_i = \langle IS_i, A, T_i, \Omega_i, O_i, R_i \rangle
\]
$IS_i$ denotes the interactive state space. This includes the physical state $S$ as well as models of the other agent $M_j$, which may be intentional or subintentional~\citep{dennett1986intentional}. In this paper, we ascribe intentional models to the other agent as they model the other agent's beliefs and capabilities as a rational agent. $A = A_i \times A_j$ is the set of joint actions of both agents. $T_i$ represents the transition function, $T_i$: $S \times A \times S \xrightarrow{} [0, 1]$. The transition function is defined over the physical states and excludes the other agent's models. This is a consequence of the model non-manipulability assumption -- an agent's actions do not directly influence the other agent's models. $\Omega_i$ is the set of agent $i$'s observations. $O$ is the observation function, $O_i$: $S \times A \times \Omega \xrightarrow{} [0, 1]$. The observation function is defined over the physical state space only as a consequence of the model non-observability assumption -- other's model parameters may not be observed directly. $R_i$ defines the reward function for agent $i$, $R_i$: $S_i \times A \xrightarrow{} \R$. The reward function for I-POMDPs usually assigns utilities to the other agent's physical states.

We limit our attention to a finitely nested I-POMDP, in which the interactive state space $IS_{i,l}$ at strategy level $l$ is defined bottom up as,
\begin{small}
\begin{equation*}
    \begin{aligned}
        IS_{i,0} & = S, & \Theta_{j,0} = \{\langle b_{j,0}, \hat{\theta}_{j} \rangle : b_{j,0} \in \Delta(IS_{j,0})\} \\
        IS_{i,1} & = S \times M_{j,0}, & \Theta_{j,1} = \{\langle b_{j,1}, \hat{\theta}_{j} \rangle : b_{j,1} \in \Delta(IS_{j,1})\} \\
        & \vdots \\
        IS_{i,l} & = S \times M_{j,l-1}, & \Theta_{j,l} = \{\langle b_{j,l}, \hat{\theta}_{j} \rangle : b_{j,l} \in \Delta(IS_{j,l})\}.
    \end{aligned}
\end{equation*}
\end{small}
Above, $\hat{\theta}_{j}$ represents agent $j$'s frame, defined as $\hat{\theta}_{j} = \langle A_{j}, \Omega_{j}, T_{j}, O_{j}, R_{j}, OC_{j} \rangle$. Here, $OC_{j}$ represents $j$'s optimality criterion and the other terms are as defined previously. $\Theta_{j}$ is the set of agent $j$'s intentional models, defined as $\theta_{j} = \langle b_{j}, \hat{\theta}_{j} \rangle$. The interactive state space is typically restricted to a finite set of $j$'s models, which are updated after every interaction to account for the belief update of agent $j$. The interactive state space for agent $i$ at level $l$ can be then defined as,
\[
         IS_{i,l} = S \times \textsf{Reach}(\Theta_{j,l-1}, H), \quad \Theta_{j,l} = \{\langle b_{j,l}, \hat{\theta}_{j} \rangle : b_{j,l} \in \Delta(IS_{j,l})\}.
\]
Here, \textsf{Reach}$(\Theta_{j,l-1}, H)$ is the set of level $l-1$ models that $j$ could have in $H$ steps; \textsf{Reach} $(\Theta_{j,l-1}, 0)$ $= \Theta_{j,l-1}$. We obtain \textsf{Reach}() by repeatedly updating $j$'s beliefs in the models in $\Theta_{j,l-1}$. 

%----------------------------------------------
\section{Modeling Cyber Deception using Factored I-POMDPs}
\label{sec:modeling}
%----------------------------------------------

Engaging and deceiving human attackers into intruding controlled systems and accessing obfuscated data offers
a proactive approach to computer and information security. It wastes attacker resources and potentially misleads
the attacker. Importantly, it offers an untapped opportunity to understand attackers' beliefs, capabilities, and preferences and how they evolve by sifting the detailed activity logs. Identifying these mental and physical states not only informs the defender about the attacker’s intent, but also guides new ways of deceiving the attacker. In this section, we first introduce our domain of cyber deception and subsequently discuss how it can be modeled in a factored I-POMDP.

%We apply I-POMDP$_\X$ to the problem of cyber deception.

%\setlength{\tabcolsep}{18pt}

%~~~~~~~~~~~~~~~~~~~~~~~~~~~~~~~~~~~~~~~~~~~~~~~~~~~~~~
\subsection{Cyber Deception Domain}
\label{subsec:domain}
%~~~~~~~~~~~~~~~~~~~~~~~~~~~~~~~~~~~~~~~~~~~~~~~~~~~~~~

% \renewcommand{\arraystretch}{1.5}
The cyber deception domain models the interactions between the attacker and the defender on a single honeypot host system. A state of the interaction is modeled using 11 state variables defining a total of 4,608 states. Table~\ref{table:domain_states} briefly summarizes the state space. The \texttt{S\_DATA\_DECOYS} and \texttt{C\_DATA\_DECOYS} state variables represent the presence of sensitive data decoys and critical data decoys. The \texttt{HOST\_HAS\_DATA} variable represents the true type of valuable data on the system. We assume that a system cannot have two different types of valuable data simultaneously. This is a reasonable assumption because usually different hosts on enterprise networks possess different assets. We differentiate between \texttt{sensitive\_data} and \texttt{critical\_data} as distinct targets. Sensitive data, for example, includes private data of employees, high ranking officials, or any data that the attacker would profit from stealing. Also, in practical scenarios, honeypots never contain any real valuable data. Consequently, in the cyber deception domain in this paper, the \texttt{HOST\_HAS\_DATA}  is \texttt{none}. However, the attacker is unaware of the honeypot or the data decoys and hence forms a belief over this state variable. Thus, the \texttt{HOST\_HAS\_DATA} variable gives a subjective view of the attacker being deceived. 

\begin{table}[!ht]
\caption{The state of the cyber deception domain is comprised of 11 variables.}
\centering
\begin{tabular}{ lll } 
\toprule
\textbf{State Variable Name} & \textbf{Values} & \textbf{Description}\\
\midrule
\texttt{PRIVS\_DECEPTION} & \texttt{user, root, none} & Deceptive reporting of privileges \\
\hline
\texttt{S\_DATA\_DECOYS} & \texttt{yes, no} & Presence of sensitive data decoys \\
\hline
\texttt{C\_DATA\_DECOYS} & \texttt{yes, no} & Presence of critical data decoys \\
\hline
\multirow{2}{*}{\texttt{HOST\_HAS\_DATA}} & \texttt{sensitive\_data,} & Type of valuable data \\
& \texttt{critical\_data, none} & on the system\\
%& \texttt{none} \\
\hline
\texttt{DATA\_ACCESS\_PRIVS} & \texttt{user, root} & Privileges required to access or find data \\
\hline
\texttt{ATTACKER\_PRIVS} & \texttt{user, root} & Attacker's highest privileges \\
\hline
\texttt{DATA\_FOUND} & \texttt{yes, no} & Valuable data found by the attacker \\
\hline
\texttt{VULN\_FOUND} & \texttt{yes, no} & Local \textit{PrivEsc} discovered by attacker\\
\hline
\texttt{IMPACT\_CAUSED} & \texttt{yes, no} & Attack successful \\
\hline
\texttt{ATTACKER\_STATUS} & \texttt{active, inactive} & Presence of attacker on the host \\
\hline
\texttt{HOST\_HAS\_VULN} & \texttt{yes, no} & Presence of local \textit{PrivEsc} vulnerability\\
\bottomrule
\end{tabular}
\label{table:domain_states}
\end{table}

There are 5 observation variables for the attacker which make a total of 48 unique observations. We include three different types of attackers; the \textit{data exfil} attacker, \textit{data manipulator} and \textit{persistent threat}. 
%These are modeled as different frames of agent $j$ in the I-POMDP$_\X$. 
The \textit{data exfil} attacker represents a threat that aims to steal valuable private data from the host. The \textit{data manipulator} attacker represents a threat that seeks to manipulate data that is critical for the operation of a business or a physical target. Thus, the  \textit{data exfil} attacker targets \texttt{sensitive\_data} in the system and the \textit{data manipulator} attacker targets \texttt{critical\_data}.  The \textit{persistent threat} attacker wants to establish a strong presence in the system at a high privilege level.

\begin{table}[!ht]
\caption{The actions available to the attacker.}
\centering
\begin{tabular}{ lll } 
\toprule
\textbf{Action name} & \textbf{States affected} & \textbf{Description} \\
\midrule
\texttt{FILE\_RECON\_SDATA} & \texttt{DATA\_FOUND} & Search for sensitive data for theft \\
\hline
\texttt{FILE\_RECON\_CDATA} & \texttt{DATA\_FOUND} & Search for critical data for manipulation \\
\hline
\texttt{VULN\_RECON} & \texttt{VULN\_FOUND} & Search for local \textit{PrivEsc} vulnerability \\
\hline
\texttt{PRIV\_ESC} & \texttt{ATTACKER\_PRIVS} & Exploit local \textit{PrivEsc} vulnerability\\
\hline
\texttt{CHECK\_ROOT} & \texttt{none} & Check availability of root privileges \\
\hline
\texttt{START\_EXFIL} & \texttt{IMPACT\_CAUSED} & Upload critical data over network \\
\hline
\texttt{PERSIST} & \texttt{IMPACT\_CAUSED} & Establish a permanent presence in the system \\
\hline
\texttt{MANIPULATE\_DATA} & \texttt{IMPACT\_CAUSED} & Manipulate stored data \\
\hline
\texttt{EXIT} & \texttt{ATTACKER\_STATUS} & Terminate the attack \\
\bottomrule
\end{tabular}
\label{table: domain_actions}
\end{table}

The attacker in the interaction can perform one of 9 actions to gather information about the system, manipulate the system, or take action on objectives. Table~\ref{table: domain_actions} briefly summarizes the actions available to the attacker. The \texttt{FILE\_RECON\_SDATA} and \texttt{FILE\_RECON\_CDATA} actions cause the \texttt{DATA\_FOUND} variable to transition to \texttt{yes}. The \texttt{FILE\_RECON\_SDATA} action is slightly worse at finding data than the \texttt{FILE\_RECON\_CDATA}. This reflects the fact that private sensitive information is slightly difficult to find because it is often stored in user directories in arbitrary locations. On the other hand, critical data, like service configuration or database files, are stored in well-known locations on the system. The attacker gets information about the \texttt{DATA\_FOUND} transition through the \texttt{DATA} observation variable. It simulates the data discovery phase of an attack. \texttt{VULN\_RECON} is another action that works similarly and causes the \texttt{VULN\_FOUND} transition to \texttt{yes}. This transition depicts the attacker looking for vulnerabilities to raise privileges. Depending on the type of the attacker, the \texttt{START\_EXFIL}, \texttt{MANIPULATE\_DATA}, or \texttt{PERSIST} actions can be performed to achieve the attacker's main objectives. We assume that the attacker is unable to discern between decoy data and real data, and hence, unable to determine which variable influences the \texttt{DATA\_FOUND} state transition during file discovery. The attacker, however, can distinguish between different types of valuable data. So, if the system contains data that is different from what the attacker expects, the attacker can observe this from the \texttt{DISCREPANCY} observation variable. As \texttt{DATA} and \texttt{DISCREPANCY} are separate observation variables, the attacker can observe a discrepancy even when data has been found. When this occurs, the attacker develops a belief over the decoy data states as the host can have only one type of data. This realistically models a situation in which the attacker encounters multiple decoys of different types and suspects deception.

% \begin{figure}
%     \centering
%     \includegraphics[width=0.6\textwidth]{images/deception_dynamics.png}
%     \caption{The dynamics of the cyber deception I-POMDP show how the defender can influence the attacker's observations indirectly by manipulating the states. The defender cannot directly manipulate the same states as the attacker.}
%     \label{fig:deception_dynamics}
% \end{figure}

The defender in the interaction starts with complete information about the system. The defender's actions mostly govern the deployment and removal of different types of decoys. These actions influence the \texttt{S\_DATA\_DECOYS} and \texttt{C\_DATA\_DECOYS} states. Additionally, the defender can influence the attacker's observations about his privileges through the \texttt{PRIVS\_DECEPTION} state. The defender gets perfect observations whenever the attacker interacts with a decoy. Additionally, the defender gets stochastic observations about the attacker's actions through the \texttt{LOG\_INFERENCE} observation variable. 
%
%Figure \ref{fig:deception_dynamics} shows the dynamics including the attacker and defender observation functions. %
The attacker is rewarded for exiting the system after causing an impact. For the \textit{data exfil} and \textit{data manipulator} attacker types, this is achieved by performing the \texttt{START\_EXFIL} and \texttt{MANIPULATE\_DATA} actions respectively. The \textit{persistent threat} attacker is rewarded for getting root level persistence in the system.

\begin{figure}[!ht]
    \centering
    \includegraphics[width=0.6\textwidth]{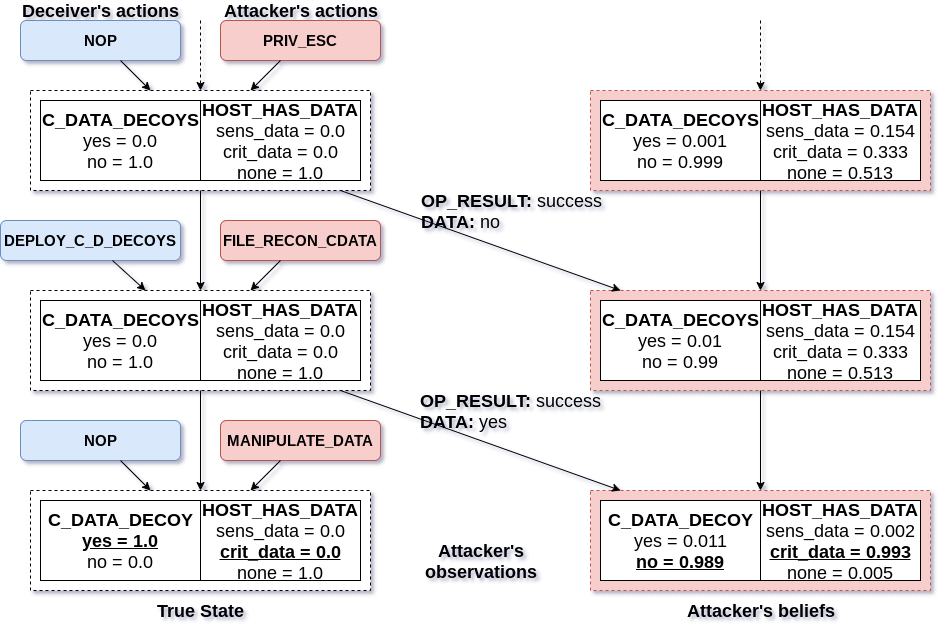}
    \caption{The attacker starts with a low prior belief on the existence of decoys and an active defender. If decoys are indistinguishable from real data, the attacker attributes his observation to the existence of real data even when the host has none.}
    \label{fig:deception_sim}
    \vspace{-0.1in}
\end{figure}

Figure~\ref{fig:deception_sim} illustrates a scenario taken from an actual simulation run with the \textit{data manipulator} attacker type. Initially, the attacker has a non-zero belief over the existence of data on the system. However, the true state of the system on the left shows that the system does not actually contain any data. In the absence of the defender or any static data decoys, the attacker will eventually update his beliefs to accurately reflect the reality by performing the \texttt{FILE\_RECON\_CDATA} action and observing the result. However, to avoid this belief state, the defender deploys data decoys when the attacker acts. The attacker's inability to tell the difference between decoy data and real data and his prior belief about the absence of decoys leads him to attribute his observations to the existence of real data leading to the attacker being deceived.

\begin{figure}[!ht]
    \centering
    \begin{subfigure}[b]{0.5\textwidth}
        \centering
        \includegraphics[width=1.0\textwidth]{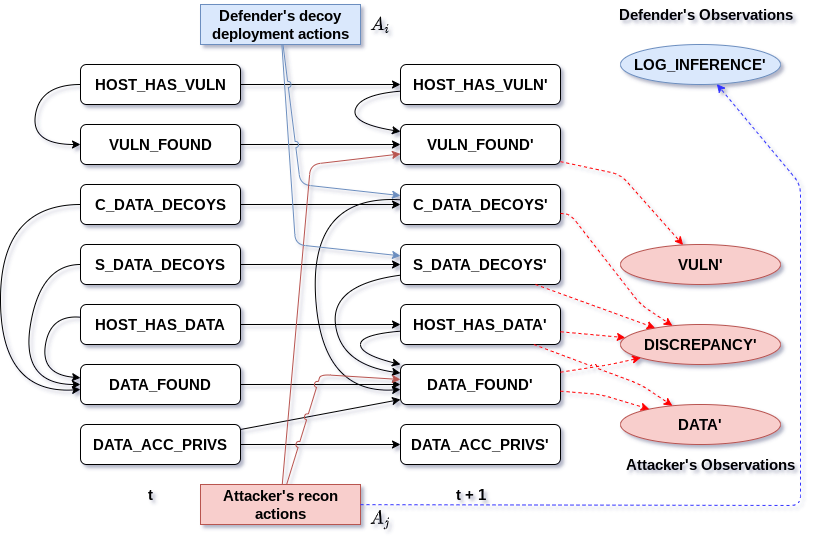}
        \caption{Dynamics compactly represented as a two time-slice DBN for select joint actions and observation variables.}
        \label{fig:deception_dynamics}
    \end{subfigure}
    \hfill
    \begin{subfigure}[b]{0.45\textwidth}
        \centering
        \includegraphics[width=1.0\textwidth]{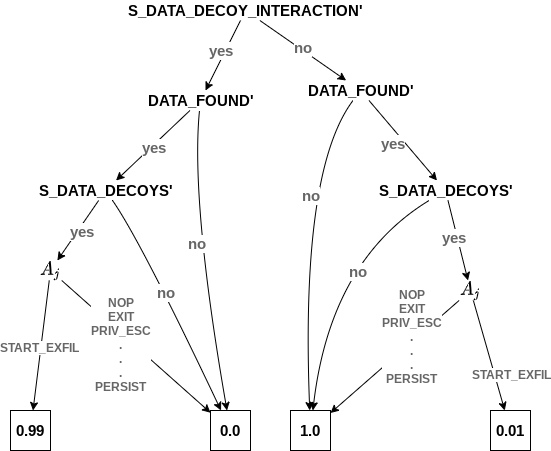}
        \caption{An ADD representing the observation function $P^{\texttt{NOP}}(\texttt{S\_DATA\_DECOY\_INTR'}|\mathcal{X}', A_{j})$}
        \label{fig:ADD}
    \end{subfigure}
    \caption{I-POMDP$_\X$ representation of the cyber deception domain.}
    \label{fig:I-POMDP_X}
    \vspace{-0.15in}
\end{figure}

%~~~~~~~~~~~~~~~~~~~~~~~~~~~~~~~~~~~~~~~~~~~~~~~~~~~~~
\subsection{Factored I-POMDPs for Modeling Cyber Deception}
\label{subsec:factored-IPOMDP}
%~~~~~~~~~~~~~~~~~~~~~~~~~~~~~~~~~~~~~~~~~~~~~~~~~~~~~

Factored POMDPs have been effective toward solving structured problems with large state and observation spaces~\citep{Feng,poupart2005exploiting}. Motivated by this observation, we extend the finitely-nested I-POMDP  reviewed in Section~\ref{sec:background} to its factored representation, I-POMDP$_\X$. Formally, this extension is defined as:
\[
    \text{I-POMDP}_\X = \langle \mathcal{IS}_i, A, T_i, \mathcal{Y}_{i}, O_{i}, \mathcal{R}_{i} \rangle
\]
$\mathcal{IS}_i$ is the factored interactive state space consisting of physical state factors $\mathcal{X}$ and agent $j$'s models $M_{j}$. In a finitely-nested I-POMDP$_\X$ the set $M_{j}$ is bounded similarly to finitely-nested I-POMDPs. Action set $A$ is defined exactly as before. We use algebraic decision diagrams (ADDs)~\citep{bahar1997algebric} to represent the factors for agent $i$'s transition, observation, and reward functions compactly. $T_{i}$ defines the transition function represented using ADDs as $P^{a_{i},a_{j}}(\mathcal{X}'|\mathcal{X})$ for $a_{i} \in A_{i}$ and $a_{j} \in A_{j}$. $\mathcal{Y}_{i}$ is the set of observation variables which make up the observation space. $O_{i}$ is the observation function represented as ADDs, $P^{a_{i},a_{j}}(\mathcal{Y}'_{i}|\mathcal{X}')$. $\mathcal{R}_{i}$ defines the reward function for agent $i$. The reward function is also represented as an ADD,  $\mathcal{R}^{a_{i},a_{j}}(\mathcal{X})$.

We illustrate I-POMDP$_\X$ by modeling the cyber deception domain of Section~\ref{subsec:domain} in the framework.  Figure~\ref{fig:deception_dynamics} shows the DBN for select state and observation variables given that the attacker engages in reconnaissance actions. The two slices in the DBN represent the sets of pre- and post-action state variables, $\mathcal{X} = \{X_{1}, ..., X_{n}\}$ and $\mathcal{X}' = \{X'_{1}, ..., X'_{n}\}$ where $X_{n}$ represents a single state variable. Similarly, $\mathcal{Y}'_{i} = \{Y'_{i_{1}}, ...,Y'_{i_{n}}\}$ and $\mathcal{Y}'_{j} = \{Y'_{j_{1}}, ...,Y'_{j_{n}}\}$ denote the sets of observation variables for agents $i$ and $j$ respectively. The ADD $P^{a_{i}}(\mathcal{X}'|\mathcal{X}, A_{j}) = P^{a_{i}}(X_{1}'|X_2', \ldots, X_n',\mathcal{X}, A_{j}) \times ... \times P^{a_{i}}(X_{n}'|\mathcal{X}, A_{j})$ represents the complete transition function for action $A_{i} = a_{i}$. This is analogous to the \textit{complete action diagram} defined by Hoey et al.~\citep{hoey2013spudd} for MDPs. Similarly, the observation function is represented using the ADD (Fig.~\ref{fig:ADD}), $P^{a_{i}}(\mathcal{Y}'_{i}|\mathcal{X}', A_{j}) = P^{a_{i}}(Y'_{i_{1}}|\mathcal{X}', A_{j}) \times ... \times P^{a_{i}}(Y'_{i_{n}}|\mathcal{X}', A_{j})$  which is analogous to the \textit{complete observation diagram}~\citep{Feng}. Additionally, in an I-POMDP$_\X$, agent $i$ also recursively updates the beliefs of agent $j$. The attacker types are modeled as frames in $M_j$. Let $M_{j} = \{m_{j_{1}}:\langle b_{j_{1}}, \hat{\theta}_{j_{1}} \rangle, ..., m_{j_{n}}:\langle b_{j_{q}}, \hat{\theta}_{j_{r}} \rangle\}$ be the set of all models in $\textsf{Reach}(\Theta_{j,l-1}, H)$. Because neither $a_{j}$ nor $o_{j}$ are directly accessible to agent $i$, they are represented as ADDs $P(A_{j}|M_{j})$ and $P^{a_{i}}(\mathcal{Y}'_{j}|\mathcal{X}',A_{j})$. The distribution over $M_{j}'$ is then $P^{a_{i}}(M_{j}'|M_{j}, \mathcal{Y}'_{j}, A_{j}, \mathcal{X}') = P^{a_{i}}(M_{j}'|M_{j},A_{j},\mathcal{Y}'_{j}) \times P^{a_{i}}(\mathcal{Y}'_{j}|\mathcal{X}',A_{j})$. Using these factors, we can now define the distribution over $\mathcal{X}'$ and $M_{j}'$ given action $a_{i}$ and observation $o_{i}$ as a single ADD using existential abstraction: 
\begin{equation}
\begin{aligned}
    P^{a_{i},o_{i}}(M_{j}',\mathcal{X}'| M_{j},\mathcal{X}) 
    %& = \sum_{A_{j}, \mathcal{Y}'_{j}} P^{a_{i},o_{i}}(\mathcal{X}',M_{j}',A_{j}, \mathcal{Y}_{j}'|\mathcal{X}, M_{j}) \nonumber\\
    & = \sum\nolimits_{A_{j}, \mathcal{Y}'_{j}} P^{a_{i},o_{i}}(\mathcal{Y}_{j}', M_{j}',\mathcal{X}',A_{j}| M_{j},\mathcal{X})\\
    & = \sum_{A_{j}, \mathcal{Y}'_{j}} P^{a_{i}}(\mathcal{X}'|\mathcal{X}, A_{j}) P^{a_{i}}(\mathcal{Y}'_{i}|\mathcal{X}', A_{j}) P(A_{j}|M_{j}) P^{a_{i}}(M_{j}'| M_{j}, A_{j}, \mathcal{Y}'_{j}, \mathcal{X}'). 
\end{aligned}
\label{eqn:joint_trans}
\end{equation}
Here, the ADD $P^{a_{i}}(\mathcal{X}'|\mathcal{X}, A_{j})$ compactly represents  $T_{i}(s^{t-1},a^{t-1}_{i},a^{t-1}_{j},s^{t})$,  $P^{a_{i}}(\mathcal{Y}'_{i}|\mathcal{X}', A_{j})$ represents the probabilities $O_{i}(s^{t},a^{t-1}_{i},a^{t-1}_{j},o^{t}_{i})$, $P(A_{j}|M_{j})$ represents $P(a^{t-1}_{j}|\theta_{j}^{t-1})$, and $P^{a_{i}}(M_{j}'| M_{j}, A_{j}, \mathcal{Y}'_{j}, \mathcal{X}')$ represents the recursive belief update transitions $\tau_{\theta^{t}_{j}}(b^{t-1}_{j}, a^{t-1}_{j}, o^{t}_{j}, b^{t}_{j}) O_{j}(s^{t},a^{t-1}_{i},a^{t-1}_{j},o_{j}^{t})$ of the original I-POMDP. Thus, the constructed ADD $P^{a_{i},o_{i}}(\mathcal{X}',M_{j}'|\mathcal{X}  M_{j})$ contains the transition probabilities for all interactive state variables given action $a_{i}$ and observation $o_{i}$. The I-POMDP$_\X$ belief update can then be computed as:
\begin{align}
    b_i^{a_{i},o_{i}}(\mathcal{X}', M_{j}') = \sum_{\mathcal{X}, M_{j}} b(\mathcal{X}, M_{j}) \times P^{a_{i},o_{i}}(\mathcal{X}',M_{j}'|\mathcal{X}, M_{j})
\label{eqn:bu}
\end{align}
where the ADD $P^{a_{i},o_{i}}(\mathcal{X}',M_{j}'|\mathcal{X}, M_{j})$ is obtained as in Eq.~\ref{eqn:joint_trans}.

Symbolic Perseus~\citep{poupart2005exploiting} offers a relatively scalable {\em point-based} approximation technique that exploits the ADD structure of factored POMDPs. Toward generalizing this technique for I-POMDP$_\X$, we are aided by the existence of point-based value iteration for I-POMDPs~\citep{doshi2008generalized}. Subsequently, we may generalize the $\alpha$-vectors and its backup from the latter to the factored representation of I-POMDP$_\X$:
\begin{equation}
    \begin{aligned}
        \Gamma^{a_{i},*} \xleftarrow{} \alpha^{a_{i},*}(\mathcal{X},M_{j}) & = \sum_{A_{j}}R^{a_{i}}(\mathcal{X}, A_{j}) P(A_{j}|M_{j}) \\
        \Gamma^{a_{i},o_{i}} \xleftarrow{\cup} \alpha^{a_{i},o_{i}}(\mathcal{X},M_{j}) & = \gamma \sum_{\mathcal{X}', M_{j}'}P^{a_{i}, o_{i}}(\mathcal{X}',M_{j}'|\mathcal{X}, M_{j}) \alpha^{t+1}(\mathcal{X}', M_{j}'), \quad \forall \alpha^{t+1} \in \mathcal{V}^{t+1} \\
        \Gamma^{a_{i}} \xleftarrow{} \Gamma^{a_{i}, *} & \oplus_{o_{i}} \argmax_{\Gamma^{a_{i}, o_{i}}}(\alpha^{a_{i},o_{i}} \cdot b_{i}), \quad \mathcal{V}^{t} \xleftarrow{} \argmax_{\alpha^{t} \in \bigcup_{a_{i}}\Gamma^{a_i}}(\alpha^{t} \cdot b_{i}),\quad \forall b_{i} \in B_{i}
    \end{aligned}
    \label{eqn:alpha-vectors}
\end{equation}
Here, $\mathcal{V}^{t+1}$ is the set of $\alpha$-vectors from the next time step and $b_i$ is a belief point from the set of considered beliefs $B_i$. A popular way of building $B_i$ is to project an initial set of beliefs points forwards for $H$ time steps using the belief update of Eq.~\ref{eqn:bu}.

% In our work, we extend the symbolic Perseus factored POMDP solver to I-POMDP$_\X$\cite{poupart2005exploiting}. The symbolic Perseus solver has several enhancements such as cached ADD computations, and ADD approximations making it efficient for large POMDPs. We extend the POMDP belief update and PBVI backup to I-POMDP$_\X$ interactive belief update and I-PBVI backups. 

%----------------------------------------------
\section{Experiments and Analysis}
\label{sec:experiments}
%----------------------------------------------

We modeled the full cyber deception domain described in Section~\ref{subsec:domain} from the perspective of a level-1 defender using the I-POMDP$_\X{}$ framework. We implemented the generalized Symbolic Perseus  using the point-based updates of the $\alpha$-vectors and the belief set projection as given in Section~\ref{subsec:factored-IPOMDP}, in order to solve I-POMDP$_\X{}$. The solver has several enhancements such as cached ADD computations and ADD approximations for additional speed up.

We evaluate the deception policy generated by I-POMDP$_\X$ in simulations and on an actual system consisting of a standalone attacker programmed via Metasploit~\citep{maynor2011metasploit} and a defender workstation. We simulate each attacker type using the optimal policy computed by the level-0 attacker POMDP. We show these policies for each type of attacker in the supplementary material. For the simulations, we randomly sample the frame and the starting privileges of the attacker to simulate a threat with unknown intentions and privileges. The defender begins knowing about the existence of decoys on the system. The attacker, on the other hand, does not have prior knowledge about any vulnerabilities or data on the system. The defender engages with the attacker by deploying decoys, facilitating deceptive observations, or adding known vulnerabilities to the system. In the simulations, the state transitions and observations for both agents are generated by sampling from the joint transition functions and individual observation functions.

% Figures \ref{fig:dm_policy} and \ref{fig:pt_policy} show the optimal policy for two of the attacker types that we model. 
% \begin{figure}[ht]
%     \centering
%     \begin{subfigure}[b]{0.45\textwidth}
%         \centering
%         \includegraphics[width=\textwidth]{images/attacker_data_manipulator.png}
%         \caption{The optimal plan for the \textit{data manipulator} type attacker who is unaware of the defender}
%         \label{fig:dm_policy}
%     \end{subfigure}
%     \hfill
%     \begin{subfigure}[b]{0.45\textwidth}
%         \centering
%         \includegraphics[width=0.7\textwidth]{images/attacker_pt.png}
%         \caption{The optimal plan for \textit{persistent threat} type attacker}
%         \label{fig:pt_policy}
%     \end{subfigure}
%     \caption{Attacker models}
%     \label{fig: entropy_graph}
% \end{figure}
\vspace{-0.1in}
\paragraph{Simulations} We compare the I-POMDP$_\X$ policy against other passive baselines: one that does not engage and passively observes the attacker, and another which uses deception indiscriminately having deployed both sensitive and critical data decoys and all vulnerabilities in the honeypot at the beginning. We label the first baseline as NO-OP(no decoy) and the second baseline as NO-OP(all decoys). We perform the simulations for 30 trials with an attacker type randomly picked in each trial. During each trial, the defender begins not knowing the type of the attacker and believes that the state is that the attacker's privileges are not known. We set $H$ in $\textsf{Reach}(\Theta_{j,l-1},H)$ to 5. The generalized Symbolic Perseus is then run on 200 projected belief points until convergence to obtain the policy, which prescribes the subsequent actions for the defender until the end of the trial. It converges in about 6 minutes with a mean time per backup of 37 secs on Ubuntu 18 with Intel i7 and 64 GB RAM.

The NO-OP(no decoy) and NO-OP(all decoy) yielded a mean ($\pm$ std err.) of 4.30 $\pm$ 0.16 and 3.26 $\pm$ 0.20 steps of engagement with the attacker, respectively. The longest engagement among these consisted of 7 and 5 steps, respectively. With NO-OP(no decoy), the attacker spends time searching for data and attempting to escalate his privileges but without much success, finally exiting the system. With NO-OP(all decoys), the attacker either quickly exploits the vulnerabilities or encounters the data decoys but quickly exits often due to the encountered data not being as expected. However, the I-POMDP$_\X$ agent engaged with the attacker for a mean duration of 5.90 $\pm$ 0.24 with the longest interaction happening for 9 steps. It leverages the information gained by the first few observations to avoid using decoys that the attacker would find suspicious. For example, the defender first manipulates the attacker's observations about his privileges. This increases her chances of observing file enumeration or vulnerability discovery activity, forming a belief over the frames. Subsequently, the defender baits the attacker using decoys and observes the interaction to solidify her belief. This minimizes the risk of the attacker encountering unexpected decoys or noticing discrepancies.
% If the attacker does not interact with the decoys within the next few interactions, the defender removes these decoys and deploys sensitive data decoys instead. The defender is careful about the initial use critical data decoys and only takes this action when the \textit{data manipulator} attacker is looking for critical data and \textit{data exfil} attacker is likely looking for vulnerabilities.
% This minimizes the risk of the attacker encountering unexpected decoys or noticing discrepancies.???  

The simulations above are predicated on the level-1 defender believing that none of the level-0 attacker types are aware of the deception, which is the typical case. However, if the defender believes that the attacker believes that there is a small chance at 0.1 of decoys being used, we observed that the attacker often quickly exited the system as one would expect. 

\begin{figure}[!ht]
    \centering
    \begin{subfigure}[b]{0.45\textwidth}
        \centering
        \includegraphics[width=1.1\textwidth]{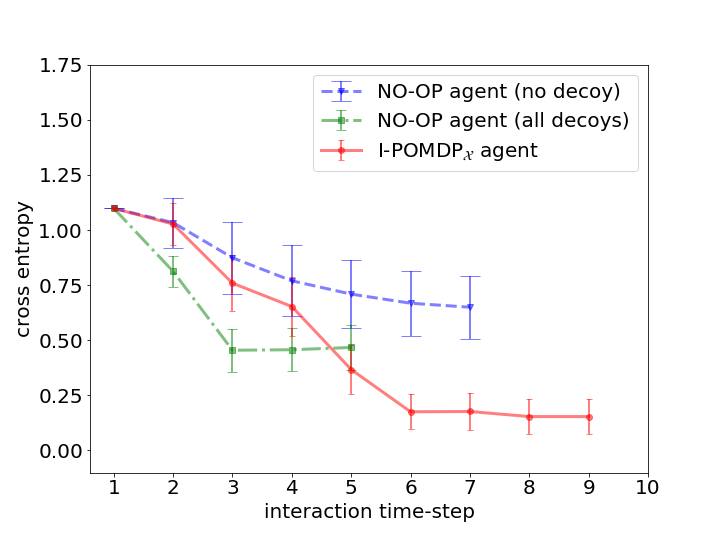}
        \caption{When engaging a defender-unaware attacker, I-POMDP$_\X$-based defender outperforms other passive agents in engaging the attackers and recognizing their intent}
        \label{fig:graph_sims}
    \end{subfigure}
    \hfill
    \begin{subfigure}[b]{0.45\textwidth}
        \centering
        \includegraphics[width=1.1\textwidth]{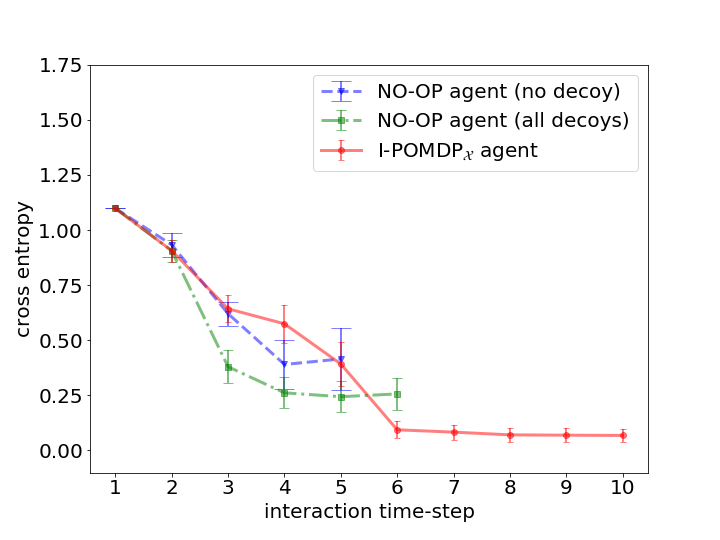}
        \caption{On the actual host deployment, the I-POMDP$_\X$-based agent uses implemented deception techniques to engage with the attacker for longer duration than other agents}
        \label{fig:actual_sims}
    \end{subfigure}
    \caption{Cross entropy (KL divergence) of the beliefs of the I-POMDP$_\X{}$ agent and other baselines in simulations. Cross entropies near zero signify good intent recognition.}
    \label{fig: entropy_graph}
    \vspace{-0.1in}
\end{figure}

% Table \ref{table: exp_duration} shows the mean duration of the engagements with each defender strategy. We see that the I-POMDP$_\X$ policy outperforms other strategies in keeping the attacker engaged. The NO-OP agent with no decoys deployed can observe the attacker. However, he is unable to extend the interaction and keep the attacker engaged. In contrast, when all decoys are deployed, the attacker encounters the wrong type of decoys and prefers not to interact further. The I-POMDP$_\X$ based strategy leverages the information gained by the first few observations to avoid using decoys that the attacker would find suspicious. As a consequence of engaging with the attacker and making observations about the system and the attacker's actions, the defender forms increasingly accurate beliefs about the attacker's frame. 

Do the extended engagements facilitated by the I-POMDP$_\X{}$ agent help in intent recognition? Figure~\ref{fig:graph_sims}(a) shows the cross-entropy between the defender's belief of the attacker's frame and the attacker's true type, as it varies across the steps of the interaction. The defender's I-POMDP$_\X{}$ policy eventually yields the lowest cross-entropy values compared to the baselines, often reaching zero in 6 steps. We show the cross-entropy for more steps because the attacker remains in the system performing a few more actions. The sharp decrease in cross-entropy in the first three steps is because the attacker's decoy interactions (if the attacker is of type {\em data exfil} or {\em manipulator}) are perfectly observed by the defender (some other interactions generate noisy observations). Multiple consecutive data reconnaissance steps filter out the persistence attacker type, and the final step of either exfiltrating the data or manipulating it allows the defender to distinguish between the remaining two attacker types. But, for the NO-OP(no decoy) with no deception, the only source of information about the attacker is his general actions, which is noisy. Hence, such a defender is unable to form accurate beliefs before the attacker leaves the interaction. For the NO-OP(all decoy) agent that indiscriminately uses deception, observations from decoy interactions are perfect, but the risk of the attacker encountering contradicting decoys and suspecting deception is also high leading to early exits.

% \begin{table}[ht!]
% \caption{Interaction duration}
% \label{table: exp_duration}
% \centering
% \begin{tabular}{ llll } 
% \toprule
% \textbf{Agent} & \textbf{Mean duration} & \textbf{Variance} & \textbf{Max}\\
% & (no.of steps) & & \textbf{duration} \\
% \midrule
% I-POMDP$_\X$ agent & 5.81 & 2.9 & 9\\
% \hline
% NO-OP agent & \multirow{2}{*}{4.42} & \multirow{2}{*}{0.61} & \multirow{2}{*}{7}\\
% (no decoys) & & \\
% \hline
% NO-OP agent & \multirow{2}{*}{2.93} & \multirow{2}{*}{1.23}  & \multirow{2}{*}{6}\\
% (all decoys) & & \\
% % Random policy. & don't know & 2.0 \\
% \bottomrule
% \end{tabular}
% \end{table}

% This shows that the I-POMDP$_\X$ policy is more effective at intent recognition than all other simpler strategies.

\begin{figure}[!ht]
    \centering
    \includegraphics[width=0.7\textwidth]{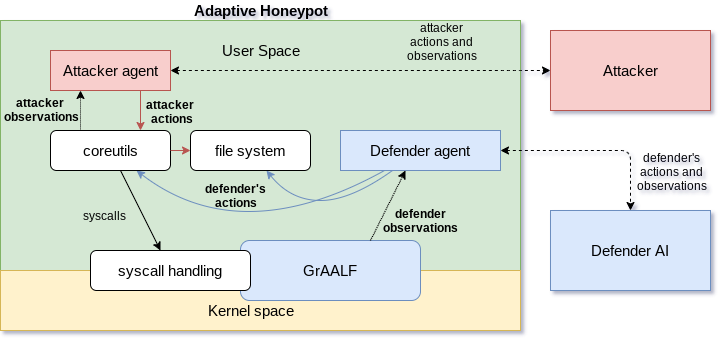}
    \caption{System architecture of the testbed used to deploy the agents. The defender manipulates the system through decoys and commonly used {\em coreutils} binaries to give deviant observations.}
    \label{fig:sys_arch}
    \vspace{-0.1in}
\end{figure}

\vspace{-0.1in}
\paragraph{Host deployment} In our next phase of experimentation, we evaluated the real-world feasibility of deploying an operational I-POMDP$_\X{}$ on a host system and testing its efficacy. The testbed consists of 3 separate hosts: the \textit{attacker}, the \textit{adaptive honeypot} and the \textit{defender}. Figure~\ref{fig:sys_arch} shows the overall architecture of our testbed implementation. The \textit{attacker} system runs a Kali Linux distribution which is well known for the variety of offensive and defensive cybersecurity tools that are preinstalled on it. The \textit{adaptive honeypot} on which the interaction takes place runs a Metasploitable 3 Linux distribution. This distribution has a wide range of builtin vulnerabilities and is commonly used to simulate victim workstations in cyber attack simulations. The \textit{adaptive honeypot} also contains an attacker agent that executes the attacks and communicates with the \textit{attacker}. The \textit{attacker agent} implements the actions given by the attacker's optimal plan located on the attacker host using realistic techniques commonly used by real attackers. We implement real exploits to facilitate privilege escalation on the host. The \textit{adaptive honeypot} also has a defender agent that implements the defender's actions and gets observations. 

The defender AI located on the \textit{defender} workstation solves the I-POMDP$_\X{}$ and computes the optimal action. For implementing the observation function, the I-POMDP$_\X$ agent monitors and analyzes the system logs to get information about the attacker's actions (i.e., observations). To enable this, we use GrAALF~\citep{setayeshfar2019graalf}, a graphical framework for processing and querying system call logs. GrAALF analyzes system call logs in real-time and provides the stochastic \texttt{LOG\_INFERENCE} observation variable values (pertaining to file and vulnerability searches) as well as the perfectly observed \texttt{DATA\_DECOY\_INTERACTION} variable values to the defender. 

Our results in Fig.~\ref{fig: entropy_graph}(b) show the adaptive deception strategy employed by the I-POMDP$_\X$ agent is better at engaging adversaries on a honeypot as compared to the passive strategies that are commonly used. While the cross entropy does not reach zero due to the challenge of accurately inferring the attacker's actions from the logs (leading to noisier observations), it gets close to zero, which is indicative of accurate intent recognition.      

%----------------------------------------------------
\section{Related Work}
\label{sec:related}
%----------------------------------------------------

AI methods are beginning to be explored for use in cyber deception. 
%In this section, we mention some other papers and research work that aims to achieve similar goals or uses related techniques. We further highlight how our contributions are unique.
An area of significant recent interest has been game-theoretic multi-agent modeling of cyber deception, which contrasts with the decision-theoretic modeling adopted in this paper.

Schlenker et al.~\citep{Schlenker2018} introduced cyber deception games based on Stackelberg games~\citep{Simaan73:Stackelberg}. These model deception during the network reconnaissance phase when the attacker is deceived into intruding a honeypot. Another similar  approach~\citep{durkota2015approximate} allocates honeypots in a network using a Stackelberg game. The game uses attack graphs to model the attacker and creates an optimal honeypot allocation strategy to lure attackers. Jajodia et al.~\citep{jajodia2017probabilistic} develop probabilistic logic to model deception during network scanning. While these efforts focus on static deployment of deception strategies at the network level, we seek active deception at the host level -- once the attacker has entered the honeypot. Further, we model individual phases of the attack in greater detail, which allows us to employ realistic deception techniques at each phase. 

At the host level, Carroll et al.~\citep{Carroll2011} models deception as a signaling game while Horak et al.~\citep{horak2017manipulating} creates a model for active deception using partially observable stochastic games. However, both of these take a high-level view modeling defender actions rather abstractly. In contrast, our defender actions are realistic and can be implemented on honeypots as demonstrated in Section~\ref{sec:experiments}. Ferguson-Walter et al.~\citep{ferguson2019game} model possible differences between the attacker's and defender's perceptions toward the interaction by modeling cyber deception as a hypergame~\citep{kovach2015hypergame}. Hypergames model different views of the game being played from the perspective of the players. While this approach similar to ours represents  the attacker's perspective of the game, we explicitly model the adversary using a subjective decision-theoretic approach and do not solve for equilibrium.

%-----------------------------------------
\section{Conclusion}
\label{sec:conclusion}
%-----------------------------------------

Our approach of utilizing automated decision making for deception to recognize attacker intent is a novel application of AI and decision making in cyber security. It elevates the extant security methods from anomaly and threat detection to intent recognition. We introduced a factored variant of the well-known I-POMDP framework, which exploits the environment structure and utilized it to model the new  cyber deception domain. Our experiments revealed that the I-POMDP$_\X{}$-based agent succeeds in engaging various types of attackers for a longer duration than passive honeypot strategies, which facilities intent recognition. Importantly, the agent is practical on a real system with logging capabilities paving the way for its deployment in actual honeypots.

\newpage

\section*{Broader Impact}

On a broader scale, the I-POMDP$_\X$ framework that we introduce makes I-POMDPs tractable to be applied to larger problems. I-POMDPs are suitable for modeling multi-agent interactions due to their ability to model opponents from the perspective of an individual. This has a multitude of applications like negotiations, studying human behavior, cognition, etc. Through our work, we hope to make I-POMDPs tractable to be applied to such domains. Another area that we hope to motivate through our research is deception in human interactions. Modeling other agents explicitly will help understand how deceptive or real information influences an individual's beliefs. This has a wide range of potential applications such as studying how biases can be exploited, the effect of fake news on individuals, and how individuals can detect deception. We hope our research will eventually motivate further research in areas like counter deception and deception resilience in agents.

At an application level, our work aims to motivate the use of AI and decision making to create informed cyber defense strategies. Our work provides a new perspective different from the traditional action-reaction dynamic that has defined interactions between cyber attackers and defenders for years. Our framework models the opponent's mental states and preferences. This will aid security teams in understanding threats at a deeper level. We hope our framework will motivate the development of adaptive and intelligent deceptive solutions that can study and predict attackers at a deeper level. Understanding attackers' mental models, inherent biases, and preferences will go a long way in forming flexible cyber defense strategies that can adapt to different threats.

% \begin{ack}
% Use unnumbered first level headings for the acknowledgments. All acknowledgments
% go at the end of the paper before the list of references. Moreover, you are required to declare 
% funding (financial activities supporting the submitted work) and competing interests (related financial activities outside the submitted work). 
% More information about this disclosure can be found at: \url{https://neurips.cc/Conferences/2020/PaperInformation/FundingDisclosure}.

% Do {\bf not} include this section in the anonymized submission, only in the final paper. You can use the \texttt{ack} environment provided in the style file to autmoatically hide this section in the anonymized submission.
% \end{ack}

\bibliographystyle{plain}
\bibliography{refs}

\begin{thebibliography}{21}
\providecommand{\natexlab}[1]{#1}
\providecommand{\url}[1]{\texttt{#1}}
\expandafter\ifx\csname urlstyle\endcsname\relax
  \providecommand{\doi}[1]{doi: #1}\else
  \providecommand{\doi}{doi: \begingroup \urlstyle{rm}\Url}\fi

\bibitem[Bahar et~al.(1997)Bahar, Frohm, Gaona, Hachtel, Macii, Pardo, and
  Somenzi]{bahar1997algebric}
R.~I. Bahar, E.~A. Frohm, C.~M. Gaona, G.~D. Hachtel, E.~Macii, A.~Pardo, and
  F.~Somenzi.
\newblock Algebric decision diagrams and their applications.
\newblock \emph{Formal methods in system design}, 10\penalty0 (2-3):\penalty0
  171--206, 1997.

\bibitem[Carroll and Grosu(2011)]{Carroll2011}
T.~E. Carroll and D.~Grosu.
\newblock {A game theoretic investigation of deception in network security}.
\newblock \emph{Security and Communication Networks}, 4\penalty0 (10):\penalty0
  1162--1172, 2011.
\newblock ISSN 19390122.
\newblock \doi{10.1002/sec.242}.

\bibitem[Dennett(1986)]{dennett1986intentional}
D.~Dennett.
\newblock Intentional systems. brainstorms, 1986.

\bibitem[Doshi(2012)]{Doshi12:Decision}
P.~Doshi.
\newblock Decision making in complex multiagent settings: A tale of two
  frameworks.
\newblock \emph{AI Magazine}, 33\penalty0 (4):\penalty0 82--95, 2012.

\bibitem[Doshi and Perez(2008)]{doshi2008generalized}
P.~Doshi and D.~Perez.
\newblock Generalized point based value iteration for interactive pomdps.
\newblock In \emph{AAAI}, pages 63--68, 2008.

\bibitem[Durkota et~al.(2015)Durkota, Lis{\`y}, Bo{\v{s}}ansk{\`y}, and
  Kiekintveld]{durkota2015approximate}
K.~Durkota, V.~Lis{\`y}, B.~Bo{\v{s}}ansk{\`y}, and C.~Kiekintveld.
\newblock Approximate solutions for attack graph games with imperfect
  information.
\newblock In \emph{International Conference on Decision and Game Theory for
  Security}, pages 228--249. Springer, 2015.

\bibitem[Feng and Hansen(2014)]{Feng}
Z.~Feng and E.~A. Hansen.
\newblock Approximate planning for factored pomdps.
\newblock In \emph{Sixth European Conference on Planning}, 2014.

\bibitem[Ferguson-Walter et~al.(2019)Ferguson-Walter, Fugate, Mauger, and
  Major]{ferguson2019game}
K.~Ferguson-Walter, S.~Fugate, J.~Mauger, and M.~Major.
\newblock Game theory for adaptive defensive cyber deception.
\newblock In \emph{Proceedings of the 6th Annual Symposium on Hot Topics in the
  Science of Security}, page~4. ACM, 2019.

\bibitem[Gmytrasiewicz and Doshi(2005)]{gmytrasiewicz2005framework}
P.~J. Gmytrasiewicz and P.~Doshi.
\newblock A framework for sequential planning in multi-agent settings.
\newblock \emph{Journal of Artificial Intelligence Research}, 24:\penalty0
  49--79, 2005.

\bibitem[Hor{\'a}k et~al.(2017)Hor{\'a}k, Zhu, and
  Bo{\v{s}}ansk{\`y}]{horak2017manipulating}
K.~Hor{\'a}k, Q.~Zhu, and B.~Bo{\v{s}}ansk{\`y}.
\newblock Manipulating adversary’s belief: A dynamic game approach to
  deception by design for proactive network security.
\newblock In \emph{International Conference on Decision and Game Theory for
  Security}, pages 273--294. Springer, 2017.

\bibitem[Jajodia et~al.(2016)Jajodia, Subrahmanian, Swarup, and
  Wang]{jajodia2016cyber}
S.~Jajodia, V.~Subrahmanian, V.~Swarup, and C.~Wang.
\newblock \emph{Cyber deception}.
\newblock Springer, 2016.

\bibitem[Jajodia et~al.(2017)Jajodia, Park, Pierazzi, Pugliese, Serra, Simari,
  and Subrahmanian]{jajodia2017probabilistic}
S.~Jajodia, N.~Park, F.~Pierazzi, A.~Pugliese, E.~Serra, G.~I. Simari, and
  V.~Subrahmanian.
\newblock A probabilistic logic of cyber deception.
\newblock \emph{IEEE Transactions on Information Forensics and Security},
  12\penalty0 (11):\penalty0 2532--2544, 2017.

\bibitem[Jesse~Hoey et~al.(1999)Jesse~Hoey, Aubin, and
  Boutilier]{hoey2013spudd}
A.~Jesse~Hoey, R.~S. Aubin, and C.~Boutilier.
\newblock Spudd: stochastic planning using decision diagrams.
\newblock \emph{Proceedings of Uncertainty in Artificial Intelligence (UAI).
  Stockholm, Sweden. Page (s)}, 15, 1999.

\bibitem[Kovach et~al.(2015)Kovach, Gibson, and Lamont]{kovach2015hypergame}
N.~S. Kovach, A.~S. Gibson, and G.~B. Lamont.
\newblock Hypergame theory: a model for conflict, misperception, and deception.
\newblock \emph{Game Theory}, 2015, 2015.

\bibitem[Maynor(2011)]{maynor2011metasploit}
D.~Maynor.
\newblock \emph{Metasploit toolkit for penetration testing, exploit
  development, and vulnerability research}.
\newblock Elsevier, 2011.

\bibitem[Pingree(2018)]{Pingree2018}
Pingree.
\newblock {Emerging Technology Analysis : Deception Techniques and Technologies
  Create Security Technology Business Opportunities}.
\newblock \emph{Trapx security}, pages 1--18, 2018.
\newblock \doi{G0027834}.

\bibitem[Poupart(2005)]{poupart2005exploiting}
P.~Poupart.
\newblock \emph{Exploiting structure to efficiently solve large scale partially
  observable Markov decision processes}.
\newblock PhD thesis, University of Toronto, 2005.

\bibitem[Schlenker et~al.(2018)Schlenker, Thakoor, Xu, Tran-Thanh, Fang,
  Vayanos, Tambe, and Vorobeychik]{Schlenker2018}
A.~Schlenker, O.~Thakoor, H.~Xu, L.~Tran-Thanh, F.~Fang, P.~Vayanos, M.~Tambe,
  and Y.~Vorobeychik.
\newblock {Deceiving cyber adversaries: A game theoretic approach}.
\newblock \emph{Proceedings of the International Joint Conference on Autonomous
  Agents and Multiagent Systems, AAMAS}, 2:\penalty0 892--900, 2018.
\newblock ISSN 15582914.

\bibitem[Setayeshfar et~al.(2019)Setayeshfar, Adkins, Jones, Lee, and
  Doshi]{setayeshfar2019graalf}
O.~Setayeshfar, C.~Adkins, M.~Jones, K.~H. Lee, and P.~Doshi.
\newblock Graalf: Supporting graphical analysis of audit logs for forensics.
\newblock \emph{arXiv preprint arXiv:1909.00902}, 2019.

\bibitem[Simaan and J.B.~Cruz(1973)]{Simaan73:Stackelberg}
M.~Simaan and J.~J.B.~Cruz.
\newblock On the stackelberg strategy in nonzero-sum games.
\newblock \emph{Journal of Optimization Theory and Applications}, 11\penalty0
  (5):\penalty0 533--555, 1973.

\bibitem[Spitzner(2003)]{Spitzner2003}
L.~Spitzner.
\newblock The honeynet project: Trapping the hackers.
\newblock \emph{IEEE Security \& Privacy}, 1\penalty0 (2):\penalty0 15--23,
  2003.

\end{thebibliography}

% \section*{References}

% References follow the acknowledgments. Use unnumbered first-level heading for
% the references. Any choice of citation style is acceptable as long as you are
% consistent. It is permissible to reduce the font size to \verb+small+ (9 point)
% when listing the references.
% {\bf Note that the Reference section does not count towards the eight pages of content that are allowed.}
% \medskip

% \small

% [1] Alexander, J.A.\ \& Mozer, M.C.\ (1995) Template-based algorithms for
% connectionist rule extraction. In G.\ Tesauro, D.S.\ Touretzky and T.K.\ Leen
% (eds.), {\it Advances in Neural Information Processing Systems 7},
% pp.\ 609--616. Cambridge, MA: MIT Press.

% [2] Bower, J.M.\ \& Beeman, D.\ (1995) {\it The Book of GENESIS: Exploring
%   Realistic Neural Models with the GEneral NEural SImulation System.}  New York:
% TELOS/Springer--Verlag.

% [3] Hasselmo, M.E., Schnell, E.\ \& Barkai, E.\ (1995) Dynamics of learning and
% recall at excitatory recurrent synapses and cholinergic modulation in rat
% hippocampal region CA3. {\it Journal of Neuroscience} {\bf 15}(7):5249-5262.

%-----------------------------
\section*{Appendix}
%-----------------------------

\subsection*{Attacker Policies}

We model the attackers using optimal policies of their level-0 POMDPs. For our problem, we define three distinct types of attackers which are modeled as separate frames in the I-POMDP. Below we discuss the optimal policies for each type.

\begin{figure}[ht!]
    \centering
    \includegraphics[width=0.7\textwidth]{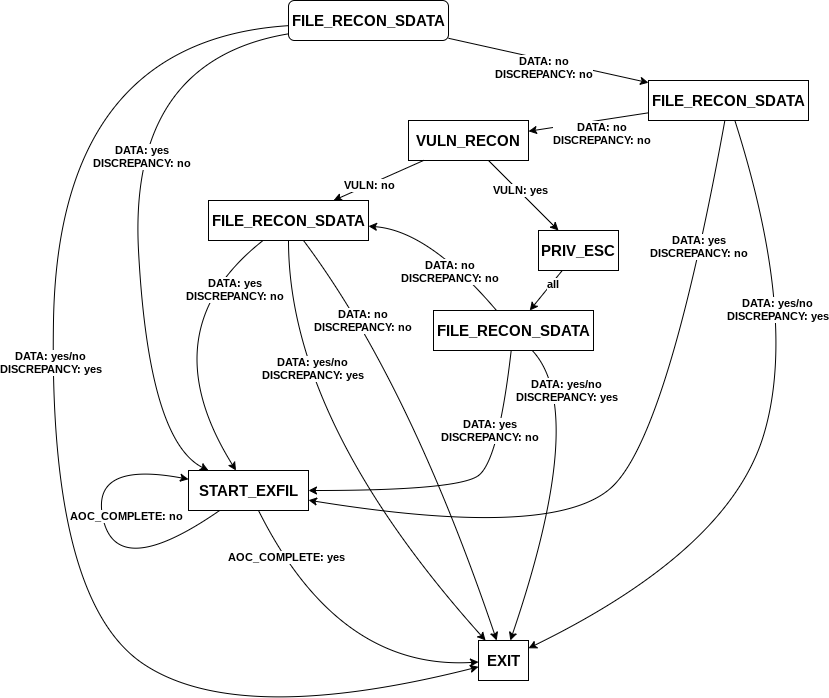}
    \caption{Optimal policy for \textit{data exfil} type attacker}
    \label{fig:de}
\end{figure}

\subsubsection*{The \textit{data exfil} attacker frame}

The \textit{data exfil} type attacker is rewarded for stealing \texttt{sensitive\_data} on the host. We model this type based on threats that steal private data and other sensitive data from systems. The attacker starts with no knowledge of the existence of data on the system. We see that the optimal policy recommends the \texttt{FILE\_RECON\_SDATA} action which simulates sensitive data discovery on computers. After failing to find data after the first few attempts, the attacker attempts to escalate privileges and search again. If the attacker encounters unexpected types of decoys, the attacker leaves since there is no reward for stealing data that is not sensitive. Also, the observation of discrepancies when data is found will inform the attacker about the possibility of deception. This is because the system only contains a single type of data. On being alerted to the possibility of being deceived, the attacker leaves the system.

\subsubsection*{The \textit{data manipulator} attacker frame}

\begin{figure}[ht!]
    \centering
    \includegraphics[width=0.6\textwidth]{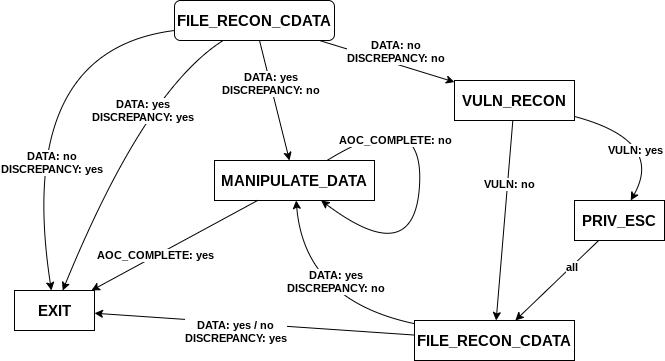}
    \caption{Optimal policy for \textit{data manipulator} type attacker}
    \label{fig:dm}
\end{figure}

The \textit{data manipulator} type attacker is rewarded for manipulating \texttt{critical\_data} on the host. This attacker type is modeled after attackers that intrude systems to manipulate data that is critical for a business operation. Similar to the \textit{data exfil} type, the attacker starts with no information about the data. The optimal policy for this attacker type recommends \texttt{FILE\_RECON\_CDATA} action in the initial steps. Because critical data like service configurations or databases are usually stored in well-known locations, the \texttt{FILE\_RECON\_CDATA} is modeled to find \texttt{critical\_data} quickly as compared to sensitive data. In the subsequent interaction steps, the attacker escalates privileges to continue the search if data is not found in the initial steps. Like the \textit{data exfil} attacker, the \textit{data manipulator} also leaves the system on observing discrepancies, suspecting deception, or on failure to find data.

\subsubsection*{The \textit{persistent threat} attacker frame}

\begin{figure}[ht!]
    \centering
    \includegraphics[width=0.4\textwidth]{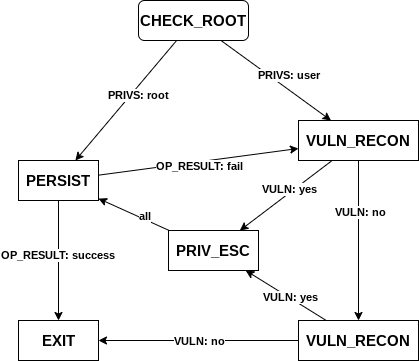}
    \caption{Optimal policy for \textit{persistent threat} type attacker}
    \label{fig:pt}
\end{figure}

The \textit{persistent threat} type attacker aims to establish root level persistence on the host. Such attacks are common. Attackers establish a strong presence in an organization's network and stay dormant for an extended duration. For this attacker type, the policy consists of vulnerability discovery actions in the initial steps. The attacker escalates privileges by performing the \texttt{PRIV\_ESC} action on finding vulnerabilities. Once the attacker has the required privileges, the \texttt{PERSIST} action is performed to complete the objective.

While all three attacker policies may seem significantly different from their actions, the defender's observations of these actions are noisy. The errors in observation come from the noisy nature of real-time log analysis. For example, the \texttt{VULN\_RECON} action models vulnerability discovery on a host. This action involves looking through the local file system for any vulnerable scripts, enumerating system information, listing services, etc. Thus a \texttt{VULN\_RECON} can be mistaken for a \texttt{FILE\_RECON\_CDATA} or a \texttt{FILE\_RECON\_SDATA} in real-time log analysis. Similarly, it is difficult to tell the difference between the \texttt{FILE\_RECON\_CDATA} and \texttt{FILE\_RECON\_SDATA} from logs alone. Hence, without baiting the attacker into performing further actions, it is challenging to infer the intent of the attacker from the first few actions.

\end{document}